\documentclass[runningheads]{llncs}
\usepackage[T1]{fontenc}
\usepackage{amssymb}
\usepackage{amsmath}
\usepackage{graphicx}
\usepackage{microtype}
\begin{document}

\title{UniStream: Multi-Expert Residual Vector Quantization for 48 kHz Causal Streaming Audio Coding}
\titlerunning{UniStream: Causal Streaming Audio Coding}
\author{Mingyu Zhao \and Zhiyong Wu\thanks{Corresponding author: zywu@sz.tsinghua.edu.cn.}}
\authorrunning{M. Zhao and Z. Wu}
\institute{Tsinghua Shenzhen International Graduate School, Tsinghua University, Shenzhen, China\\
\email{zmy24@mails.tsinghua.edu.cn, zywu@sz.tsinghua.edu.cn}}
\maketitle            
\begin{abstract}
We present UniStream, a fully causal 48\,kHz neural audio codec for streaming speech, music, and environmental sounds. At its core is Multi-Expert Residual Vector Quantization (ME-RVQ), which replaces the single shared codebook in each residual quantization layer with four expert codebooks controlled by a deterministic Top-$K$ router. Because routing decisions are derived solely from previously decoded quantized states, the decoder can reproduce the selected experts without transmitting expert identifiers, thereby expanding quantization capacity while adding 5.5M parameters. We further introduce an auxiliary Optimal Transport Conditional Flow Matching (OT-CFM) objective to regularize the quantized latent space during training. The flow module is removed entirely at inference and therefore incurs no runtime overhead. UniStream supports a 12\,kbps Top-1 mode and a 22.5\,kbps Top-2 mode within a causal 48\,kHz encoder--decoder framework, while achieving real-time GPU inference. To complement narrow-band speech metrics, we report 48\,kHz ViSQOL audio mode, ViSQOL speech mode, standard VGGish-FAD, DNSMOS P.835, and higher-rate reference comparisons with Opus and EnCodec. At 12\,kbps, UniStream-Top1 achieves PESQ and UTMOS scores comparable to EnCodec while reducing speech Mel-D from 13.07 to 8.21. At 22.5\,kbps, UniStream-Top2 achieves a ViSQOL speech-mode score of 4.67 and an environmental audio-mode score of 3.96, exceeding all evaluated systems operating at 12\,kbps or below in the latter setting. It also comes within 0.03 MOS-LQO of Opus at 24\,kbps on speech in ViSQOL audio mode. Ablation studies confirm that ME-RVQ is the primary source of quality improvement, whereas OT-CFM provides perceptual gains on speech with a mild trade-off in spectral distortion.


\end{abstract}

\section{Introduction}

Neural audio codecs have become a fundamental component of modern speech and audio systems, enabling downstream applications such as speech language modeling, real-time communication, and generative audio modeling~\cite{audiolm,valle}. SoundStream~\cite{soundstream} established the residual vector quantization (RVQ)-based encoder--quantizer--decoder paradigm, which was subsequently extended toward high-fidelity and more general audio coding by systems such as EnCodec~\cite{encodec}, DAC~\cite{dac}, SNAC~\cite{snac}, Mimi~\cite{moshi}, WavTokenizer~\cite{wavtokenizer}, and Gull~\cite{gull}. Despite these advances, designing a codec that is simultaneously causal, efficient, and robust across speech, music, and environmental audio remains challenging.

A central challenge lies in the quantization bottleneck. Most existing RVQ-based codecs assign a single shared codebook to each quantization layer, requiring the same representation space to model heterogeneous acoustic structures across multiple domains. Although effective for general reconstruction, this design limits domain-adaptive specialization, particularly because speech formants, musical harmonics, and environmental textures exhibit substantially different temporal and spectral characteristics. Mixture-of-experts (MoE) models offer a natural way to expand representational capacity with limited computational overhead~\cite{switchtransformer,stmoe}. However, existing MoE-based codec designs either place experts outside the quantization bottleneck or require additional routing information to be transmitted, increasing signaling overhead and complicating streaming deployment~\cite{switchcodec,languagecodec}.

Flow-based codecs such as FlowMAC~\cite{flowmac} and FlowDec~\cite{flowdec} improve perceptual reconstruction through conditional flow matching, but require iterative inference. This motivates using flow matching solely as a training-time regularizer for streaming codecs.

In this paper, we present \textbf{UniStream}, a fully causal 48\,kHz neural audio codec for streaming speech, music, and environmental audio. Its core component, \textbf{Multi-Expert Residual Vector Quantization (ME-RVQ)}, replaces the single shared codebook in each acoustic RVQ layer with multiple expert codebooks controlled by a learned Top-$K$ router. Because routing decisions depend only on previously decoded quantized states, the decoder can deterministically reproduce the selected experts without transmitting expert identifiers. This design expands quantization capacity while adding 5.5M parameters.

We further introduce an auxiliary \textbf{Optimal Transport Conditional Flow Matching (OT-CFM)} objective to regularize the quantized latent space during training. Unlike flow-based decoders that require iterative inference, the OT-CFM module is removed entirely after training and therefore incurs no runtime overhead. The resulting codec supports a 12\,kbps Top-1 mode for bandwidth-constrained streaming and a 22.5\,kbps Top-2 mode for higher-fidelity reconstruction.

We evaluate UniStream using conventional and full-band perceptual metrics across speech, music, and environmental audio. The results show that ME-RVQ is the primary source of improvement: UniStream-Top1 provides a competitive 12\,kbps operating point, while UniStream-Top2 approaches higher-rate references under full-band perceptual evaluation.

The main contributions of this work are summarized as follows:
\begin{itemize}
\item We propose ME-RVQ, a multi-expert quantization mechanism that expands RVQ capacity by adding 5.5M parameters and requires no expert-ID signaling.
\item We introduce a training-only OT-CFM objective that provides an additional latent-space training signal without increasing inference-time cost.
\item We develop a fully causal 48\,kHz streaming codec supporting both 12\,kbps and 22.5\,kbps operating modes.
\item We conduct a comprehensive evaluation using full-band ViSQOL, VGGish-FAD, DNSMOS P.835, and 24\,kbps Opus/EnCodec reference comparisons, addressing the limitations of narrow-band speech-only evaluation.
\end{itemize}

\section{Related Work}

\subsection{Neural Audio Codecs}

SoundStream~\cite{soundstream} established the RVQ-based encoder--quantizer--decoder framework for low-latency neural audio compression, while EnCodec~\cite{encodec} and DAC~\cite{dac} advanced high-fidelity reconstruction through adversarial and improved RVQGAN training. More recent systems, including SNAC~\cite{snac}, Mimi~\cite{moshi}, WavTokenizer~\cite{wavtokenizer}, and Gull~\cite{gull}, explore multi-scale quantization, semantic--acoustic representations, compact tokenization, and flexible sample-rate coding. However, many existing systems either rely on non-causal components, focus primarily on speech, or employ shared quantization structures that may not fully capture the heterogeneity of speech, music, and environmental audio. Recent studies have explored complementary forms of codec
adaptation. SPG-Codec~\cite{spgcodec} investigates frozen semantic
priors and bitrate-aware regulation for ultra-low-bitrate neural
speech coding, while BAMU~\cite{bamu} performs post-hoc frame-wise
RVQ depth allocation under an exact serialized-bitstream budget.
In contrast, UniStream modifies the quantization architecture
itself for causal 48\,kHz general-audio coding and enables the
decoder to reproduce expert routing without transmitting expert
identifiers.

\subsection{Quantization and Mixture-of-Experts Modeling}

Mixture-of-experts (MoE) architectures offer a natural way to increase representational capacity while activating only a subset of parameters for each input~\cite{switchtransformer,stmoe}. Recent audio coding studies have begun to incorporate domain-adaptive or expert-based mechanisms. SwitchCodec~\cite{switchcodec} explores residual-expert sparse quantization, while UniCodec~\cite{unicodec} introduces domain-adaptive modeling for unified audio coding. Language-Codec~\cite{languagecodec} modifies codec representations to better support speech language modeling. These studies demonstrate the potential of expert specialization for discrete audio representation. Nevertheless, prior residual-expert quantization methods may require explicit routing information in the bitstream. Unlike SwitchCodec, ME-RVQ derives routing solely from the cumulative quantized reconstruction available identically at the encoder and decoder. The decoder can therefore reproduce the selected experts and routing weights without transmitting expert identifiers. Unlike post-hoc bitrate allocation methods such as BAMU~\cite{bamu},
which operate on a frozen codec after training, UniStream integrates
adaptive routing directly into the causal quantization architecture.

\subsection{Flow-Based Audio Coding and Training Regularization}

Generative flow models have recently been introduced into audio coding to improve perceptual reconstruction. FlowMAC~\cite{flowmac} applies conditional flow matching to low-bitrate audio coding, while FlowDec~\cite{flowdec} extends flow-based reconstruction to full-band general audio. These methods demonstrate that flow models can improve perceptual quality by modeling complex reconstruction distributions beyond those captured by deterministic decoders. However, when a flow model is integrated into the inference-time decoder, it typically requires iterative sampling or multiple neural function evaluations, increasing computational cost and making real-time streaming more challenging.

UniStream adopts flow matching in a different manner. Rather than using a flow model as an inference-time decoder, we employ an auxiliary Optimal Transport Conditional Flow Matching (OT-CFM) objective only during training. The flow module provides an additional learning signal that encourages the quantized latent representation to retain information about the continuous encoder output. After training, the module is removed entirely, leaving the deployed codec as a fully causal encoder--quantizer--decoder system. This design preserves the regularization benefits of flow matching without introducing runtime overhead.

\section{Proposed Method}

\subsection{System Overview}

\begin{figure}[t]
\centering
\includegraphics[width=\linewidth]{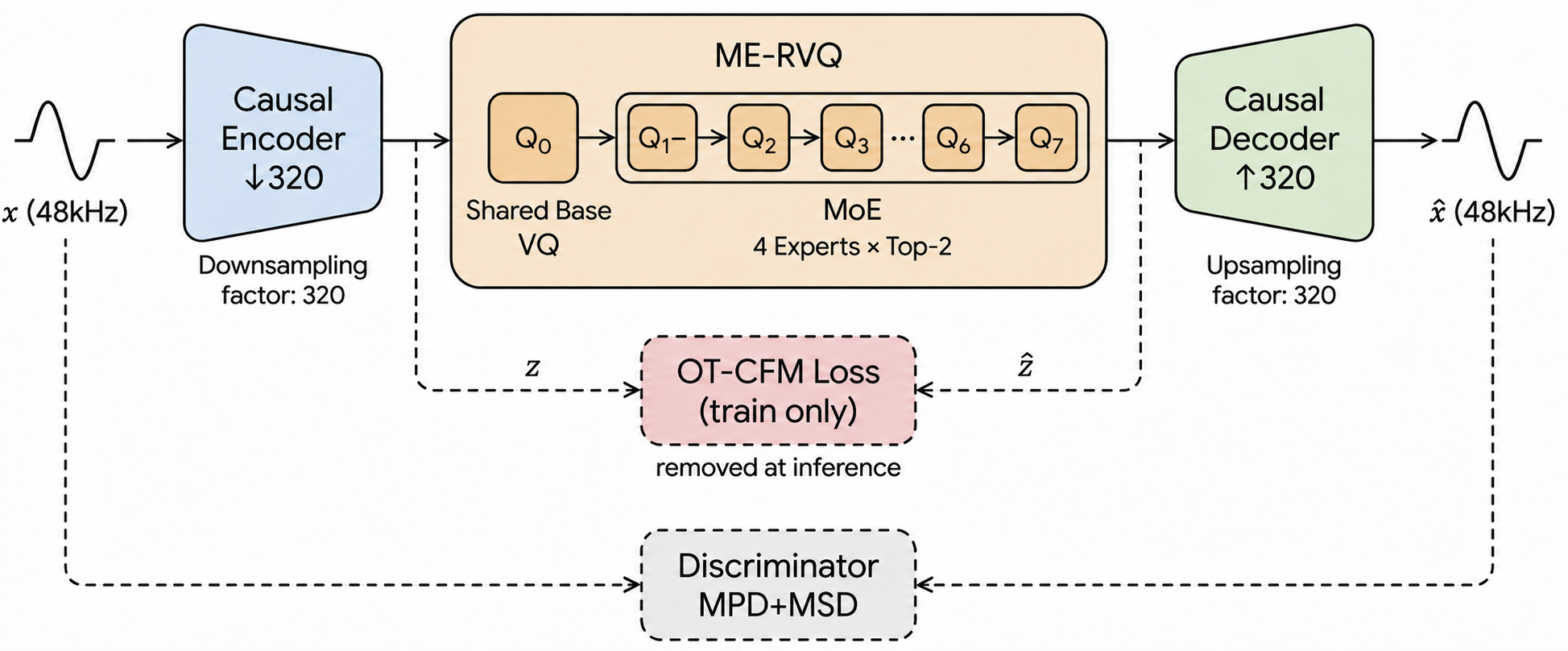}
\caption{Overview of the UniStream architecture. Dashed boxes denote training-only components; only the causal encoder, ME-RVQ quantizer, and causal decoder are retained at inference.}
\label{fig:framework}
\end{figure}

As shown in Fig.~\ref{fig:framework}, UniStream consists of a causal convolutional encoder $\mathcal{E}$, a Multi-Expert Residual Vector Quantization (ME-RVQ) module $\mathcal{Q}$, a causal convolutional decoder $\mathcal{D}$, and an auxiliary Optimal Transport Conditional Flow Matching (OT-CFM) module $\mathcal{F}$ used only during training. Given an input waveform $\mathbf{x}\in\mathbb{R}^{1\times T}$ sampled at 48\,kHz, the encoder first maps the waveform to a continuous latent representation, which is subsequently quantized and decoded to reconstruct the waveform:
\begin{equation}
\mathbf{z}=\mathcal{E}(\mathbf{x}),\qquad
\hat{\mathbf{z}}=\mathcal{Q}(\mathbf{z}),\qquad
\hat{\mathbf{x}}=\mathcal{D}(\hat{\mathbf{z}}).
\label{eq:pipeline}
\end{equation}
Here, $\mathbf{z}$ and $\hat{\mathbf{z}}$ denote the continuous and quantized latent sequences, respectively. The encoder and decoder adopt a symmetric causal CNN architecture inspired by SEANet~\cite{seanet}. Four strided convolutional blocks with strides $[2,4,5,8]$ yield a total downsampling factor of $320$, corresponding to a latent frame rate of 150\,Hz for 48\,kHz audio. Each block employs dilated causal convolutions with dilation factors $[1,3,9]$, GELU activations, and weight normalization~\cite{weightnorm}, following common practice in neural waveform generation~\cite{hifigan}. The latent dimension is set to $D=256$, and all modules used at inference are fully causal.

\subsection{Multi-Expert Residual Vector Quantization}

Standard residual vector quantization assigns a single shared codebook to each quantization layer. Although compact and efficient, this design requires one representation space to model heterogeneous acoustic structures across domains such as speech, music, and environmental audio. UniStream addresses this limitation through ME-RVQ, which equips each acoustic RVQ layer with multiple expert codebooks and a learned routing mechanism.

\subsubsection{Split RVQ Structure}

ME-RVQ comprises $L=8$ quantization layers. The first layer, $Q_0$, is a shared base vector quantization layer with $1024$ entries. The remaining layers, $Q_1$ through $Q_7$, are multi-expert residual quantization layers. These layers progressively refine the reconstruction residual using expert-specific codebooks, thereby increasing quantization capacity without altering the overall causal encoder--decoder architecture.

\subsubsection{MoE Quantizer Layer}

For each MoE quantization layer $Q_\ell$, where $\ell=1,\ldots,7$, we employ $E=4$ expert codebooks,
$\{C_\ell^{(e)}\}_{e=1}^{E}$,
each containing $1024$ entries. The router $\mathcal{R}_\ell$ takes the running reconstruction $\tilde{\mathbf{z}}_\ell$ as input and predicts frame-level expert weights:
\begin{equation}
\mathbf{g}_\ell
=
\mathrm{Softmax}\left(
\mathrm{GELU}\left(\tilde{\mathbf{z}}_\ell W_1\right)W_2
\right),
\qquad
\mathbf{g}_\ell\in\mathbb{R}^{T'\times E},
\label{eq:router}
\end{equation}
where $W_1\in\mathbb{R}^{D\times H}$, $W_2\in\mathbb{R}^{H\times E}$, and $H=128$. The softmax operation is applied along the expert dimension. For each frame, the router selects the Top-$K$ experts. The default high-quality mode uses $K=2$. Let $\mathcal{S}_{\ell,t}$ denote the set of selected experts for frame $t$. The corresponding routing weights are renormalized as
\begin{equation}
\bar{g}_{\ell,t}^{(e)}
=
\frac{g_{\ell,t}^{(e)}}
{\sum_{e'\in\mathcal{S}_{\ell,t}}g_{\ell,t}^{(e')}},
\qquad
e\in\mathcal{S}_{\ell,t}.
\label{eq:renorm}
\end{equation}
Each selected expert independently quantizes the residual
\begin{equation}
\mathbf{r}_\ell
=
\mathbf{z}-\tilde{\mathbf{z}}_\ell,
\end{equation}
and the output at frame $t$ is obtained as a weighted combination of the selected expert quantization results:
\begin{equation}
\hat{\mathbf{r}}_{\ell,t}
=
\sum_{e\in\mathcal{S}_{\ell,t}}
\bar{g}_{\ell,t}^{(e)}
\cdot
\mathrm{VQ}_{C_\ell^{(e)}}\left(\mathbf{r}_{\ell,t}\right).
\label{eq:moe_quant}
\end{equation}
Gradients are propagated through the discrete codebook lookup using the straight-through estimator~\cite{vqvae}. The running reconstruction is then updated as
\begin{equation}
\tilde{\mathbf{z}}_{\ell+1}
=
\tilde{\mathbf{z}}_\ell
+
\mathrm{sg}\left(\hat{\mathbf{r}}_\ell\right),
\label{eq:running_recon}
\end{equation}
where $\mathrm{sg}(\cdot)$ denotes the stop-gradient operation. All codebooks are updated using exponential moving average (EMA) statistics with a decay factor of $0.99$ and Laplace smoothing. The average entry utilization across all expert codebooks is 87.3\%, indicating broad usage of the available codebook entries.

\subsubsection{Routing Regularization}

To mitigate expert collapse, we employ the load-balancing loss introduced in sparse MoE models~\cite{switchtransformer}:
\begin{equation}
\mathcal{L}_{\mathrm{bal}}
=
E\sum_{e=1}^{E} f_e P_e,
\label{eq:balance}
\end{equation}
where $f_e$ denotes the fraction of tokens routed to expert $e$, and $P_e$ is the mean routing probability assigned to that expert. We further apply the router z-loss~\cite{stmoe} to stabilize the scale of the routing logits:
\begin{equation}
\mathcal{L}_{z}
=
\frac{1}{BT'}
\sum_{b,t}
\left(
\log\sum_{e}\exp\left(l_{b,t}^{(e)}\right)
\right)^2,
\label{eq:zloss}
\end{equation}
where $l_{b,t}^{(e)}$ denotes the pre-softmax routing logit for expert $e$ at batch index $b$ and frame index $t$. Both routing losses are computed in float32 for numerical stability.

\subsubsection{Operating Modes and Deterministic Decoding}

ME-RVQ supports two operating modes. In \textbf{Top-1 mode}, only the highest-ranked expert is selected in each MoE layer. The first shared layer transmits one 10-bit code index, while each of the seven MoE layers transmits one additional 10-bit code index. The resulting bitrate is
\begin{equation}
(1\times 10 + 7\times 10)\times 150
=
12\,\mathrm{kbps}.
\end{equation}
This mode targets bandwidth-constrained real-time communication.

In \textbf{Top-2 mode}, each MoE layer transmits two 10-bit code indices, resulting in
\begin{equation}
(1\times 10 + 7\times 20)\times 150
=
22.5\,\mathrm{kbps}.
\end{equation}
This mode provides higher reconstruction quality while preserving causal streaming.

A key property of ME-RVQ is deterministic decoding. At layer $\ell$, the router input is the running reconstruction
\begin{equation}
\tilde{\mathbf{z}}_\ell
=
\sum_{j<\ell}\mathbf{q}_j,
\end{equation}
which depends only on previously decoded code indices already available from the bitstream. The decoder can therefore reproduce the same expert selections and routing weights as the encoder without transmitting expert identifiers. This eliminates the expert-ID signaling overhead required by prior residual-expert quantization designs~\cite{switchcodec}. In our implementation, the router is evaluated in strict float32 precision, and bit-exact encoder--decoder agreement is verified on both an A100 GPU and a Xeon 8358 CPU.

\subsection{OT-CFM Training Regularizer}

Vector quantization inevitably discards fine-grained information when the continuous encoder output $\mathbf{z}$ is replaced by its discrete codebook reconstruction $\hat{\mathbf{z}}$. To provide an additional learning signal for the quantized latent space, we introduce an auxiliary Optimal Transport Conditional Flow Matching (OT-CFM) objective during training.

Unlike flow-based audio codecs that employ conditional flow models as iterative inference-time decoders~\cite{flowmac,flowdec}, our flow module is used exclusively during training and removed entirely at inference. Consequently, the regularizer introduces no runtime overhead or additional streaming latency.

During training, we sample
$t\sim\mathcal{U}(0,1)$
and
$\mathbf{z}_0\sim\mathcal{N}(0,I)$,
and construct the optimal-transport straight-line interpolant
\begin{equation}
\mathbf{z}_t
=
(1-t)\mathbf{z}_0+t\mathbf{z}.
\label{eq:ot_interpolant}
\end{equation}
A velocity network $v_\theta$, implemented as a 6-layer causal Transformer with a hidden dimension of $256$, $4$ attention heads, and sinusoidal time embeddings, predicts the velocity field conditioned on the quantized latent $\hat{\mathbf{z}}$. The flow-matching loss is defined as
\begin{equation}
\mathcal{L}_{\mathrm{FM}}
=
\mathbb{E}_{t,\mathbf{z}_0}
\left[
\left\|
v_\theta(\mathbf{z}_t,t,\hat{\mathbf{z}})
-
(\mathbf{z}-\mathbf{z}_0)
\right\|_2^2
\right].
\label{eq:fm_loss}
\end{equation}
Here, $\mathbf{z}-\mathbf{z}_0$ is the target velocity along the straight-line path from Gaussian noise to the encoder latent. Conditioning the velocity prediction on $\hat{\mathbf{z}}$ encourages the quantized representation to retain information about the original continuous latent. In this way, OT-CFM regularizes the latent space without introducing inference-time cost.

\subsection{Training Objective}

The generator is optimized using a combination of reconstruction, adversarial, quantization, routing, and flow-regularization losses:
\begin{equation}
\begin{aligned}
\mathcal{L}_{G}
={}&
\lambda_{\mathrm{mel}}\mathcal{L}_{\mathrm{mel}}
+\lambda_{\mathrm{stft}}\mathcal{L}_{\mathrm{stft}}
+\lambda_{\ell_1}\mathcal{L}_{\ell_1}
+\lambda_{\mathrm{adv}}\mathcal{L}_{\mathrm{adv}} \\
&+
\lambda_{\mathrm{feat}}\mathcal{L}_{\mathrm{feat}}
+\mathcal{L}_{Q}
+\lambda_{\mathrm{FM}}\mathcal{L}_{\mathrm{FM}}.
\end{aligned}
\label{eq:total_loss}
\end{equation}
The reconstruction objective comprises a multi-scale mel-spectrogram loss $\mathcal{L}_{\mathrm{mel}}$~\cite{encodec}, a single-resolution STFT magnitude loss $\mathcal{L}_{\mathrm{stft}}$~\cite{hifigan}, and a time-domain $\ell_1$ waveform loss $\mathcal{L}_{\ell_1}$. The adversarial and feature-matching losses are computed using MPD and MSD discriminators~\cite{hifigan}. Adversarial training is activated after 10k steps, allowing the generator to first learn a stable reconstruction mapping.

The quantizer loss is defined as
\begin{equation}
\mathcal{L}_{Q}
=
\mathcal{L}_{\mathrm{commit}}
+
\lambda_{\mathrm{bal}}\mathcal{L}_{\mathrm{bal}}
+
\lambda_{z}\mathcal{L}_{z},
\label{eq:quant_loss}
\end{equation}
where the commitment loss is
\begin{equation}
\mathcal{L}_{\mathrm{commit}}
=
\left\|
\mathbf{z}
-
\mathrm{sg}\left(\hat{\mathbf{z}}\right)
\right\|_2^2.
\label{eq:commit}
\end{equation}
The flow-matching loss is activated after 5k steps, once the codebooks have stabilized, to avoid disrupting early-stage quantizer learning. Based on validation performance, we set
$\lambda_{\mathrm{mel}}=45$,
$\lambda_{\mathrm{adv}}=1$,
$\lambda_{\mathrm{feat}}=2$,
$\lambda_{\mathrm{bal}}=0.01$,
$\lambda_z=0.001$,
$\lambda_{\mathrm{FM}}=1$,
$\lambda_{\ell_1}=0.1$,
and
$\lambda_{\mathrm{stft}}=1$.

\section{Experiments}

\subsection{Experimental Setup}

\textbf{Datasets and training.}
UniStream is trained on three audio domains: LibriSpeech~\cite{librispeech} for speech, MTG-Jamendo~\cite{mtgjamendo} for music, and FSD50K~\cite{fsd50k} for environmental audio. During training, the three datasets are sampled with probabilities of $0.5$, $0.3$, and $0.2$, respectively. All audio is converted to 48\,kHz mono and loudness-normalized to $-23$\,LUFS. The model is trained for 200k steps on 8 A100 GPUs with a batch size of 128 using 1-second audio segments. We use the Adam optimizer with an initial learning rate of $3\times10^{-4}$, a 1k-step warmup, and cosine learning-rate decay. For evaluation, we use fixed subsets of 200 utterances from
LibriSpeech test-clean, 100 clips from the MTG-Jamendo test set, and
94 clips from the FSD50K evaluation set. The same evaluation files
and preprocessing settings are used for all compared systems.
LibriSpeech waveforms are resampled from 16\,kHz to 48\,kHz before
codec evaluation.

\textbf{Baselines.}
We compare UniStream with representative conventional and neural audio codecs, including Opus~\cite{opus}, EnCodec~\cite{encodec}, DAC~\cite{dac}, SNAC~\cite{snac}, and Mimi in Moshi~\cite{moshi}. Opus and EnCodec are evaluated at both 12\,kbps and 24\,kbps. The 12\,kbps configurations serve as matched-rate baselines for UniStream-Top1, whereas the 24\,kbps configurations provide higher-rate references for UniStream-Top2, which operates at 22.5\,kbps. DAC, SNAC, and Mimi are evaluated using their official checkpoints or recommended configurations, and their actual operating bitrates are reported in the tables. The neural baselines are evaluated using their official checkpoints
and recommended configurations rather than being retrained on our
mixed-domain corpus. Therefore, these comparisons are intended as
practical reference comparisons under their standard settings. Non-causal systems are included as reference systems rather than direct streaming baselines. For EnCodec, we use the official non-causal 48\,kHz stereo checkpoint. The 24\,kbps setting denotes the total stereo bitrate, corresponding to 12\,kbps per channel. Mono inputs are duplicated across the two channels, and the left reconstructed channel is used for evaluation.

\textbf{Evaluation metrics.}
We report both conventional speech-codec metrics and full-band perceptual measures to evaluate 48\,kHz audio reconstruction. For speech, we include PESQ~\cite{pesq}, STOI~\cite{stoi}, UTMOS~\cite{utmos}, and DNSMOS P.835~\cite{dnsmos_p835}. We also report mel-spectral distortion (Mel-D) for speech, music, and environmental audio. Because PESQ and STOI are computed after resampling to 16\,kHz, they primarily characterize speech reconstruction and do not fully evaluate full-band 48\,kHz audio. We therefore report ViSQOL v3~\cite{visqol} in two modes: 48\,kHz audio mode for speech, music, and environmental audio, and 16\,kHz speech mode for speech-only evaluation. For distributional audio quality, we compute standard VGGish-FAD using AudioSet-pretrained VGGish embeddings~\cite{vggish,audioset,fad}, where lower values indicate greater similarity to the reference distribution. VGGish-FAD is computed after resampling the audio to 16\,kHz and is used as a standard distributional complement to, rather than a replacement for, the 48\,kHz ViSQOL evaluation.

\subsection{Conventional Objective Metrics}

Table~\ref{tab:main_conventional} reports the conventional objective results. UniStream-Top1 is rate-matched to the 12\,kbps Opus and EnCodec configurations; Opus is causal, whereas the 48\,kHz EnCodec checkpoint is non-causal. Relative to EnCodec at the same nominal bitrate, it achieves comparable PESQ and UTMOS scores while substantially reducing mel-spectral distortion, although its STOI score is lower. UniStream-Top2 operates at 22.5\,kbps and represents a higher-quality operating point; therefore, it is not directly ranked against the 12\,kbps baselines.

\begin{table}[t]
\centering
\caption{Conventional objective results. Non-causal systems are included as reference systems. UniStream-Top2 operates at 22.5\,kbps and is not directly comparable to the 12\,kbps baselines.}
\label{tab:main_conventional}
\scriptsize
\setlength{\tabcolsep}{3pt}
\begin{tabular}{lccccccc}
\hline
System & Rate & Causal & PESQ & UTMOS & STOI & Sp. Mel-D & Mus. Mel-D / Env. Mel-D \\
\hline
Opus & 12k & Yes & 3.99 & 3.81 & 0.936 & 13.76 & 22.51 / 28.52 \\
EnCodec & 12k & No & 2.88 & 3.35 & 0.878 & 13.07 & 9.20 / 13.18 \\
DAC & 8k & No & 2.95 & 3.41 & 0.883 & 6.94 & 7.81 / 8.12 \\
SNAC & 2.6k & No & 2.62 & 3.39 & 0.881 & 7.69 & 8.42 / 8.74 \\
Mimi & 1.1k & Yes & 2.71 & 3.12 & 0.854 & 9.43 & 11.37 / 12.05 \\
UniStream-T1 & 12k & Yes & 2.81 & 3.34 & 0.826 & 8.21 & 10.72 / 10.16 \\
UniStream-T2 & 22.5k & Yes & 3.41 & 3.87 & 0.864 & 7.25 & 9.54 / 7.91 \\
\hline
\end{tabular}
\end{table}

At 12\,kbps, UniStream-Top1 obtains a PESQ score of 2.81 and a UTMOS score of 3.34, both close to those of EnCodec at the same nominal bitrate. Its main advantage lies in spectral reconstruction: compared with EnCodec, speech Mel-D decreases from 13.07 to 8.21, while environmental Mel-D decreases from 13.18 to 10.16. However, UniStream-Top1 yields a lower STOI score of 0.826. This discrepancy highlights the importance of considering complementary perceptual and full-band metrics rather than relying on a single evaluation criterion.

\subsection{Full-Band ViSQOL Evaluation}

Table~\ref{tab:visqol} reports the ViSQOL results. Audio-mode scores are computed at 48\,kHz for speech, music, and environmental audio, whereas speech-mode scores are computed on speech using the 16\,kHz speech model. This evaluation complements conventional speech metrics with a perceptually motivated assessment of full-band reconstruction.

\begin{table}[t]
\centering
\caption{ViSQOL MOS-LQO results. Audio mode is evaluated at 48\,kHz on speech, music, and environmental audio, while speech mode is evaluated on speech only. Higher is better.}
\label{tab:visqol}
\scriptsize
\setlength{\tabcolsep}{5pt}
\begin{tabular}{lccccc}
\hline
System & Rate & Audio-Sp. & Audio-Mus. & Audio-Env. & Speech Mode \\
\hline
Opus & 12k & 4.32 & 2.43 & 2.43 & 4.63 \\
EnCodec & 12k & 4.03 & 4.13 & 3.91 & 4.15 \\
DAC & 8k & 4.46 & 4.10 & 3.94 & 4.63 \\
SNAC & 2.6k & 4.35 & 3.98 & 3.83 & 4.17 \\
Mimi & 1.1k & 4.03 & 2.93 & 2.88 & 4.43 \\
UniStream-T1 & 12k & 4.35 & 3.82 & 3.70 & 4.43 \\
UniStream-T2 & 22.5k & 4.42 & 4.06 & 3.96 & 4.67 \\
Opus & 24k & 4.45 & 4.23 & 4.07 & 4.77 \\
EnCodec & 24k & 4.10 & 4.22 & 4.02 & 4.31 \\
\hline
\end{tabular}
\end{table}

UniStream-Top1 achieves a speech audio-mode score of 4.35, matching SNAC while operating as a causal 12\,kbps codec. UniStream-Top2 further improves the speech audio-mode score to 4.42 and achieves 4.67 in ViSQOL speech mode, the highest score among the evaluated neural codecs operating at 22.5\,kbps or below. On environmental audio, UniStream-Top2 achieves 3.96, outperforming all listed systems operating at 12\,kbps or below. Compared with Opus at 24\,kbps, UniStream-Top2 is within 0.03 MOS-LQO on speech in ViSQOL audio mode while operating at a slightly lower bitrate of 22.5\,kbps. These results demonstrate that UniStream-Top2 is competitive with higher-rate reference codecs under full-band perceptual evaluation.

\subsection{VGGish-FAD Evaluation}

Because FAD depends on the choice of embedding encoder, we report standard VGGish-FAD using AudioSet-pretrained VGGish embeddings. The results are shown in Table~\ref{tab:vggish_fad}. VGGish-FAD measures the distributional similarity between reconstructed and reference audio, with lower values indicating closer alignment with the reference distribution.

\begin{table}[t]
\centering
\caption{VGGish-FAD results on speech, music, and environmental audio. Lower is better.}
\label{tab:vggish_fad}
\scriptsize
\setlength{\tabcolsep}{8pt}
\begin{tabular}{lccc}
\hline
System & Speech & Music & Environmental \\
\hline
Opus & 0.60 & 1.71 & 0.95 \\
EnCodec & 1.68 & 0.37 & 0.55 \\
DAC & 0.18 & 0.32 & 0.43 \\
SNAC & 0.43 & 0.40 & 0.86 \\
Mimi & 0.24 & 1.28 & 0.77 \\
UniStream-T1 & 0.72 & 1.48 & 0.95 \\
UniStream-T2 & 0.27 & 0.68 & 0.44 \\
\hline
\end{tabular}
\end{table}

UniStream-Top2 achieves a speech FAD of 0.27, outperforming Opus, EnCodec, SNAC, and UniStream-Top1, while approaching the strongest reference systems. On environmental audio, UniStream-Top2 obtains 0.44, close to DAC at 0.43 and lower than SNAC, Mimi, and Opus. Its music FAD of 0.68 remains higher than those of EnCodec, DAC, and SNAC, indicating that further improvement is needed in music-domain modeling and harmonic reconstruction.

\subsection{DNSMOS P.835 Speech Evaluation}

Table~\ref{tab:dnsmos} reports DNSMOS P.835 scores on LibriSpeech test-clean. We report SIG, BAK, and OVR, which estimate speech-signal quality, background-noise quality, and overall quality, respectively. These non-intrusive scores provide a complementary estimate of perceived speech quality.

\begin{table}[t]
\centering
\caption{DNSMOS P.835 results on speech. SIG, BAK, and OVR estimate speech-signal quality, background-noise quality, and overall quality, respectively. Higher is better.}
\label{tab:dnsmos}
\scriptsize
\setlength{\tabcolsep}{8pt}
\begin{tabular}{lcccc}
\hline
System & Rate & SIG & BAK & OVR \\
\hline
Opus & 12k & 3.56 & 3.97 & 3.24 \\
EnCodec & 12k & 3.46 & 3.78 & 3.06 \\
DAC & 8k & 3.63 & 4.01 & 3.33 \\
SNAC & 2.6k & 3.56 & 3.97 & 3.25 \\
Mimi & 1.1k & 3.58 & 4.03 & 3.29 \\
UniStream-T1 & 12k & 3.43 & 3.84 & 3.07 \\
UniStream-T2 & 22.5k & 3.55 & 3.89 & 3.20 \\
Opus & 24k & 3.60 & 4.03 & 3.31 \\
EnCodec & 24k & 3.55 & 3.90 & 3.21 \\
\hline
\end{tabular}
\end{table}

UniStream-Top2 obtains SIG, BAK, and OVR scores of 3.55, 3.89, and 3.20, respectively, closely matching EnCodec at 24\,kbps, which achieves 3.55, 3.90, and 3.21. Although DAC and Mimi achieve slightly higher DNSMOS scores, they operate under different model configurations and bitrates and are therefore included as reference systems. Overall, the DNSMOS results place UniStream-Top2 in a similar perceptual speech-quality range to the higher-rate reference codecs.

\subsection{Ablation Study}

Table~\ref{tab:ablation} examines the contributions of ME-RVQ and OT-CFM. Removing ME-RVQ decreases PESQ from 3.41 to 3.20 and UTMOS from 3.87 to 3.80, while consistently increasing mel-spectral distortion across speech, music, and environmental audio. These results confirm that ME-RVQ is the primary contributor to the overall quality improvement.

\begin{table}[t]
\centering
\caption{Ablation study at 22.5\,kbps. Lower Mel-D is better; higher PESQ, UTMOS, and STOI are better.}
\label{tab:ablation}
\scriptsize
\setlength{\tabcolsep}{6pt}
\begin{tabular}{lcccccc}
\hline
Configuration & PESQ & UTMOS & STOI & Sp. Mel-D & Mus. Mel-D & Env. Mel-D \\
\hline
Full model & 3.41 & 3.87 & 0.864 & 7.25 & 9.54 & 7.91 \\
w/o ME-RVQ & 3.20 & 3.80 & 0.856 & 7.72 & 10.19 & 8.36 \\
w/o OT-CFM & 3.37 & 3.86 & 0.865 & 7.07 & 9.34 & 7.87 \\
\hline
\end{tabular}
\end{table}

The effect of OT-CFM is more nuanced. Removing OT-CFM slightly decreases PESQ from 3.41 to 3.37 and UTMOS from 3.87 to 3.86, while marginally improving Mel-D across all three domains. We therefore interpret OT-CFM as an auxiliary speech-perceptual regularizer rather than the primary driver of general audio reconstruction quality. This complementary behavior is desirable for the proposed system:
ME-RVQ provides the main capacity and reconstruction improvement,
whereas OT-CFM supplies an additional training signal for perceptual
speech quality without increasing inference-time complexity. This behavior is consistent with its role as an additional latent-space training objective, which is not explicitly optimized to minimize mel-spectral distortion.

\subsection{Efficiency}

UniStream contains 75.6M parameters during training. Because the OT-CFM module is discarded at inference, the deployed codec contains 47.9M parameters. ME-RVQ introduces 5.5M additional parameters. On an A100 GPU, UniStream achieves an end-to-end RTF of 0.144, comprising encoder and decoder RTFs of 0.134 and 0.010, respectively. On a Xeon 8358 CPU, the model achieves an RTF of 1.05, corresponding to near-real-time inference. With a total downsampling factor of $320$, the latent frame interval is
\begin{equation}
\frac{320}{48000}\approx 6.7\,\mathrm{ms}.
\end{equation}
This value denotes the latent-frame interval rather than the complete
end-to-end algorithmic latency. Since all inference-time modules are
causal and use no look-ahead, it provides a lower bound on the
streaming latency. The GPU RTF of 0.144 supports real-time deployment,
while the CPU RTF of 1.05 indicates approximately real-time
operation.

\section{Conclusion}

We presented UniStream, a fully causal 48\,kHz neural audio codec for speech, music, and environmental audio. Its core ME-RVQ module expands RVQ capacity through expert codebooks while enabling deterministic decoder-side routing without expert-ID signaling. A training-only OT-CFM objective provides modest perceptual gains on speech without inference-time overhead. Experiments show that ME-RVQ is the primary source of improvement: UniStream-Top1 provides a competitive 12\,kbps operating point with improved spectral reconstruction, while UniStream-Top2 approaches higher-rate references under full-band perceptual evaluation and performs strongly on environmental audio. Future work will investigate subjective listening tests, entropy coding, and deployment on mobile and edge hardware.

\bibliographystyle{splncs04}
\bibliography{references}

@article{audiolm,
author  = {Z. Borsos and R. Marinier and D. Vincent and E. Kharitonov and O. Pietquin and M. Sharifi and D. Roblek and O. Teboul and D. Grangier and M. Tagliasacchi and N. Zeghidour},
title   = {{AudioLM}: A Language Modeling Approach to Audio Generation},
journal = {IEEE/ACM Transactions on Audio, Speech, and Language Processing},
volume  = {31},
pages   = {2523--2533},
year    = {2023}
}

@article{valle,
author  = {C. Wang and S. Chen and Y. Wu and Z. Zhang and L. Zhou and S. Liu and Z. Chen and Y. Liu and H. Wang and J. Li and L. He and S. Zhao and F. Wei},
title   = {Neural Codec Language Models Are Zero-Shot Text to Speech Synthesizers},
journal = {IEEE Transactions on Audio, Speech and Language Processing},
volume  = {33},
pages   = {705--718},
year    = {2025}
}

@article{soundstream,
author  = {N. Zeghidour and A. Luebs and A. Omran and J. Skoglund and M. Tagliasacchi},
title   = {{SoundStream}: An End-to-End Neural Audio Codec},
journal = {IEEE/ACM Transactions on Audio, Speech, and Language Processing},
volume  = {30},
pages   = {495--507},
year    = {2022}
}

@article{encodec,
author  = {A. D\'{e}fossez and J. Copet and G. Synnaeve and Y. Adi},
title   = {High Fidelity Neural Audio Compression},
journal = {Transactions on Machine Learning Research},
year    = {2023}
}

@inproceedings{dac,
author    = {R. Kumar and P. Seetharaman and A. Luebs and I. Kumar and K. Kumar},
title     = {High-Fidelity Audio Compression with Improved {RVQGAN}},
booktitle = {Proc. Advances in Neural Information Processing Systems (NeurIPS)},
year      = {2023}
}

@inproceedings{seanet,
author       = {M. Tagliasacchi and Y. Li and K. Misiunas and D. Roblek},
title        = {{SEANet}: A Multi-Modal Speech Enhancement Network},
booktitle    = {Proc. Annual Conference of the International Speech Communication Association (INTERSPEECH)},
pages        = {1126--1130},
year         = {2020},
organization = {ISCA}
}

@inproceedings{snac,
author    = {H. Siuzdak and F. Gr\"{o}tschla and L. A. Lanzend\"{o}rfer},
title     = {{SNAC}: Multi-Scale Neural Audio Codec},
booktitle = {Proc. Audio Imagination: NeurIPS 2024 Workshop on AI-Driven Speech, Music, and Sound Generation},
year      = {2024}
}

@article{moshi,
author  = {Alexandre D\'{e}fossez and Laurent Mazar\'{e} and Manu Orsini and Am\'{e}lie Royer and Patrick P\'{e}rez and Herv\'{e} J\'{e}gou and Edouard Grave and Neil Zeghidour},
title   = {{Moshi}: A Speech-Text Foundation Model for Real-Time Dialogue},
journal = {arXiv preprint arXiv:2410.00037},
year    = {2024}
}

@inproceedings{wavtokenizer,
author    = {S. Ji and Z. Jiang and W. Wang and Y. Chen and M. Fang and J. Zuo and Q. Yang and X. Cheng and Z. Wang and R. Li and Z. Zhang and X. Yang and R. Huang and Y. Jiang and Q. Chen and S. Zheng and W. Wang and Z. Zhao},
title     = {{WavTokenizer}: An Efficient Acoustic Discrete Codec Tokenizer for Audio Language Modeling},
booktitle = {Proc. International Conference on Learning Representations (ICLR)},
year      = {2025}
}

@article{gull,
author  = {Y. Luo and J. Yu and H. Chen and R. Gu and C. Weng},
title   = {{Gull}: A Generative Multifunctional Audio Codec},
journal = {arXiv preprint arXiv:2404.04947},
year    = {2024}
}

@article{switchtransformer,
author  = {W. Fedus and B. Zoph and N. Shazeer},
title   = {Switch {Transformers}: Scaling to Trillion Parameter Models with Simple and Efficient Sparsity},
journal = {Journal of Machine Learning Research},
volume  = {23},
number  = {120},
pages   = {1--39},
year    = {2022}
}

@article{stmoe,
author  = {Barret Zoph and Irwan Bello and Sameer Kumar and Nan Du and Yanping Huang and Jeff Dean and Noam Shazeer and William Fedus},
title   = {{ST-MoE}: Designing Stable and Transferable Sparse Expert Models},
journal = {arXiv preprint arXiv:2202.08906},
year    = {2022}
}

@inproceedings{switchcodec,
author       = {Xiangbo Wang and Wenbin Jiang and Jin Wang and Yubo You and Sheng Fang and Fei Wen},
title        = {{SwitchCodec}: Adaptive Residual-Expert Sparse Quantization for High-Fidelity Neural Audio Coding},
booktitle    = {Proc. IEEE International Conference on Acoustics, Speech and Signal Processing (ICASSP)},
year         = {2026},
organization = {IEEE}
}

@inproceedings{unicodec,
author    = {Y. Jiang and Q. Chen and S. Ji and Y. Xi and W. Wang and C. Zhang and X. Yue and S. Zhang and H. Li},
title     = {{UniCodec}: Unified Audio Codec with Single Domain-Adaptive Codebook},
booktitle = {Proc. Annual Meeting of the Association for Computational Linguistics (ACL)},
pages     = {19112--19124},
year      = {2025}
}

@inproceedings{languagecodec,
author    = {S. Ji and M. Fang and J. Zuo and Z. Jiang and D. Wang and H. Wang and H. Huang and Z. Zhao},
title     = {{Language-Codec}: Bridging Discrete Codec Representations and Speech Language Models},
booktitle = {Proc. Annual Meeting of the Association for Computational Linguistics (ACL)},
pages     = {13332--13345},
year      = {2025}
}

@inproceedings{flowmac,
author       = {N. Pia and M. Strauss and M. Multrus and B. Edler},
title        = {{FlowMAC}: Conditional Flow Matching for Audio Coding at Low Bit Rates},
booktitle    = {Proc. IEEE International Conference on Acoustics, Speech and Signal Processing (ICASSP)},
pages        = {1--5},
year         = {2025},
organization = {IEEE}
}

@inproceedings{flowdec,
author    = {S. Welker and M. Le and R. T. Q. Chen and W.-N. Hsu and T. Gerkmann and A. Richard and Y.-C. Wu},
title     = {{FlowDec}: A Flow-Based Full-Band General Audio Codec with High Perceptual Quality},
booktitle = {Proc. International Conference on Learning Representations (ICLR)},
year      = {2025}
}

@inproceedings{weightnorm,
author    = {T. Salimans and D. P. Kingma},
title     = {Weight Normalization: A Simple Reparameterization to Accelerate Training of Deep Neural Networks},
booktitle = {Proc. Advances in Neural Information Processing Systems (NeurIPS)},
year      = {2016}
}

@inproceedings{hifigan,
author    = {J. Kong and J. Kim and J. Bae},
title     = {{HiFi-GAN}: Generative Adversarial Networks for Efficient and High Fidelity Speech Synthesis},
booktitle = {Proc. Advances in Neural Information Processing Systems (NeurIPS)},
year      = {2020}
}

@inproceedings{vqvae,
author    = {A. van den Oord and O. Vinyals and K. Kavukcuoglu},
title     = {Neural Discrete Representation Learning},
booktitle = {Proc. Advances in Neural Information Processing Systems (NeurIPS)},
year      = {2017}
}

@inproceedings{librispeech,
author       = {V. Panayotov and G. Chen and D. Povey and S. Khudanpur},
title        = {{LibriSpeech}: An {ASR} Corpus Based on Public Domain Audio Books},
booktitle    = {Proc. IEEE International Conference on Acoustics, Speech and Signal Processing (ICASSP)},
pages        = {5206--5210},
year         = {2015},
organization = {IEEE}
}

@inproceedings{mtgjamendo,
author    = {D. Bogdanov and M. Won and P. Tovstogan and A. Porter and X. Serra},
title     = {The {MTG-Jamendo} Dataset for Automatic Music Tagging},
booktitle = {Proc. Machine Learning for Music Discovery Workshop at the International Conference on Machine Learning (ICML)},
year      = {2019}
}

@article{fsd50k,
author  = {E. Fonseca and X. Favory and J. Pons and F. Font and X. Serra},
title   = {{FSD50K}: An Open Dataset of Human-Labeled Sound Events},
journal = {IEEE/ACM Transactions on Audio, Speech, and Language Processing},
volume  = {30},
pages   = {829--852},
year    = {2022}
}

@inproceedings{pesq,
author       = {A. W. Rix and J. G. Beerends and M. P. Hollier and A. P. Hekstra},
title        = {Perceptual Evaluation of Speech Quality ({PESQ}): A New Method for Speech Quality Assessment of Telephone Networks and Codecs},
booktitle    = {Proc. IEEE International Conference on Acoustics, Speech and Signal Processing (ICASSP)},
pages        = {749--752},
year         = {2001},
organization = {IEEE}
}

@inproceedings{utmos,
author       = {T. Saeki and D. Xin and W. Nakata and S. Takamichi and H. Saruwatari},
title        = {{UTMOS}: {UTokyo-SaruLab} System for {VoiceMOS} Challenge 2022},
booktitle    = {Proc. Annual Conference of the International Speech Communication Association (INTERSPEECH)},
pages        = {4521--4525},
year         = {2022},
organization = {ISCA}
}

@article{stoi,
author  = {C. H. Taal and R. C. Hendriks and R. Heusdens and J. Jensen},
title   = {An Algorithm for Intelligibility Prediction of Time--Frequency Weighted Noisy Speech},
journal = {IEEE Transactions on Audio, Speech, and Language Processing},
volume  = {19},
number  = {7},
pages   = {2125--2136},
year    = {2011}
}

@inproceedings{fad,
author       = {K. Kilgour and M. Zuluaga and D. Roblek and M. Sharifi},
title        = {{Fr\'{e}chet} Audio Distance: A Reference-Free Metric for Evaluating Music Enhancement Algorithms},
booktitle    = {Proc. Annual Conference of the International Speech Communication Association (INTERSPEECH)},
pages        = {2350--2354},
year         = {2019},
organization = {ISCA}
}

@inproceedings{visqol,
author       = {Michael Chinen and Felicia S. C. Lim and Jan Skoglund and Nikita Gureev and Feargus O'Gorman and Andrew Hines},
title        = {{ViSQOL} v3: An Open Source Production Ready Objective Speech and Audio Metric},
booktitle    = {Proc. Twelfth International Conference on Quality of Multimedia Experience (QoMEX)},
pages        = {1--6},
year         = {2020},
organization = {IEEE}
}

@inproceedings{dnsmos_p835,
author       = {Chandan K. A. Reddy and Vishak Gopal and Ross Cutler},
title        = {{DNSMOS} {P.835}: A Non-Intrusive Perceptual Objective Speech Quality Metric to Evaluate Noise Suppressors},
booktitle    = {Proc. IEEE International Conference on Acoustics, Speech and Signal Processing (ICASSP)},
pages        = {886--890},
year         = {2022},
organization = {IEEE}
}

@inproceedings{vggish,
author       = {Shawn Hershey and Sourish Chaudhuri and Daniel P. W. Ellis and Jort F. Gemmeke and Aren Jansen and R. Channing Moore and Manoj Plakal and Devin Platt and Rif A. Saurous and Bryan Seybold and Malcolm Slaney and Ron J. Weiss and Kevin Wilson},
title        = {{CNN} Architectures for Large-Scale Audio Classification},
booktitle    = {Proc. IEEE International Conference on Acoustics, Speech and Signal Processing (ICASSP)},
pages        = {131--135},
year         = {2017},
organization = {IEEE}
}

@inproceedings{audioset,
author       = {Jort F. Gemmeke and Daniel P. W. Ellis and Dylan Freedman and Aren Jansen and Wade Lawrence and R. Channing Moore and Manoj Plakal and Marvin Ritter},
title        = {{Audio Set}: An Ontology and Human-Labeled Dataset for Audio Events},
booktitle    = {Proc. IEEE International Conference on Acoustics, Speech and Signal Processing (ICASSP)},
pages        = {776--780},
year         = {2017},
organization = {IEEE}
}

@misc{opus,
author       = {Jean-Marc Valin and Koen Vos and Timothy B. Terriberry},
title        = {Definition of the {Opus} Audio Codec},
howpublished = {{RFC} 6716},
publisher    = {RFC Editor},
year         = {2012},
month        = sep,
doi          = {10.17487/RFC6716}
}

@article{spgcodec,
  author  = {Zhao, Mingyu and Lin, Zijian and Wei, Kun and Wu, Zhiyong},
  title = {{SPG-Codec}: Exploring the Role and Boundaries of Semantic Priors in Ultra-Low-Bitrate Neural Speech Coding},
  journal = {arXiv preprint arXiv:2604.26296},
  year    = {2026}
}

@article{bamu,
  author  = {Zhao, Mingyu and Lin, Zijian and Feng, Yutang and Chen, Jiatao and Wang, Fan and Luo, Jiehui and Ding, Yuhao and Zhang, Jinchao and Wu, Zhiyong},
  title = {{BAMU}: Bitstream-Aware Marginal-Utility Allocation for Frozen Pretrained Neural Speech Codecs},
  journal = {arXiv preprint arXiv:2608.08432},
  year    = {2026}
}

\end{document}